\documentclass[aps,prb,amsmath,amssymb,floatfix,twocolumn]{revtex4}
\pdfoutput=1
\usepackage{graphicx}
\usepackage{hyperref}
\usepackage{times}

\begin{document}

\title{Singlet versus triplet  particle-hole condensates in quantum oscillations in cuprates}

\author{David Garcia-Aldea}
\author{Sudip Chakravarty}
\affiliation{Department of Physics and Astronomy, University of
California Los Angeles\\ Los Angeles, CA 90095-1547}

\date{\today}

\begin{abstract}
Quantum oscillations in a tilted magnetic field offer the possibility of distinguishing singlet versus triplet order parameters in the particle hole channel provided the measurements reflect a putative ``normal'' state of a density wave obtained by applying a high magnetic field at low temperatures. A theoretical analysis is given that compares spin density wave, a singlet $d$-density wave, and a triplet $d$-density wave. While the existence of a spin zero in the  oscillation amplitude is a necessary consequence of a singlet order parameter, a triplet order parameter may or may not exhibit a spin zero, making it a quantitative issue  that depends on the actual extremal orbits on the Fermi surface. Nonetheless, a theoretical analysis can shed light on the striking recent measurements in $\mathrm{YBa_{2}Cu_{3}O_{6+\delta}}$.

\end{abstract}

\maketitle
\section{Introduction}
A remarkable experiment in 2007~\cite{Doiron-Leyraud:2007} involving quantum oscillations of the Hall coefficient in  $\mathrm{YBa_{2}Cu_{3}O_{6.5}}$ has raised an important question regarding the ground state, as the superconducting dome is crushed by a high magnetic field, high enough to destroy superconductivity to a large extent but not high enough on the scale of electronic energies.~\cite{Chakravarty:2008,Chakravarty:2008b} Subsequently a large number of such experiments have provided overall consistency, but perhaps not full agreement in all respects. Another striking observation has been a similar measurement in electron doped $\mathrm{Nd_{2-x}Ce_{x}CuO_{4}}$ where the measurements are carried out between $30-65$ Tesla, far above the upper critical field $H_{c2}$ less than $10$ Tesla.~\cite{Helm:2009} If any universality is to hold,  the mechanism of quantum oscillation must be  the same  in electron and hole doped cuprates,~\cite{Eun:2009} even though the measurements in hole doped cuprates are perhaps below $H_{c2}$, but unquestionably carried out  in the resistive state. An emerging view is that the oscillations result from Fermi pockets formed by a suitable density wave state, a condensate in the particle-hole channel. We explore this possibility to shed light on two conflicting experiments in a tilted magnetic field, one in which no spin zeros were found~\cite{Sebastian:2009,Sebastian:2010} and the other in which they were found.~\cite{Ramshaw:2010}

Condensates in particle-hole channel are fundamentally different from condensates in the particle-particle channel---superconductors. Since there is no exchange requirement between a particle and a hole, the symmetry of the orbital wave function does not constrain the symmetry of the spin wave function. A given orbital symmetry can come in both singlet and triplet varieties. These order parameters are very nicely classified on the basis of the angular momentum channel.~\cite{Nayak:2000,Nersesyan:1991}  Here, we concentrate on the $s$ and $d$-orbital channels leaving out the possible $p$-channel, which probably is unlikely for the cuprates. Thus, the singlet (sDDW) and the triplet (tDDW) $d$-density wave order parameters are, considering only the two-dimensional and the two-fold commensurate  case:
\begin{eqnarray}
\langle c^{\dagger}_{\alpha}({\bf k}+{\bf Q})
{c_\beta}({\bf k})\rangle
&=& i\Phi_{\bf Q} f({\bf k}) \delta_{\alpha\beta},\\
\langle c^{\dagger}_{\alpha}({\bf k}+{\bf Q})
{c_\beta}({\bf k})\rangle
&=& i\Phi_{\bf Q}f({\bf k}) \hat{n}\cdot \boldsymbol{\sigma}_{\alpha \beta}.
\end{eqnarray}
Here $c_{\alpha}({\bf k})$ is the fermion destruction operator of spin index $\alpha$ and the $d$-wave form factor is  
\begin{equation}
f({\bf k})=(\cos k_{x}a-\cos k_{y}a)/2.
\end{equation}
The vector ${\bf Q}=(\pi/a,\pi/a)$ and the magnitude $\Phi_{\bf Q}$ is real. The singlet order parameter transforms as identity in the spin space, while the triplet transforms as $\hat{n}\cdot \boldsymbol{\sigma}_{\alpha \beta}$, $\hat n$ being the direction of the spin quantization axis and $\boldsymbol{\sigma}$ the standard Pauli matrix. We shall assume that spin-orbit coupling is negligibly small.

 The singlet and the triplet $s$-wave density waves are the conventional charge (CDW) and spin density waves (SDW), defined by setting $f({\bf k})=1$ and removing the factor of $i$:
\begin{eqnarray}
\langle c^{\dagger}_{\alpha}({\bf k}+{\bf Q})
c_{\beta}({\bf k})\rangle
&=& \Phi_{\bf Q} \delta_{\alpha \beta},\\
\langle c^{\dagger}_{\alpha}({\bf k}+{\bf Q})
c_{\beta}({\bf k})\rangle
&=& \Phi_{\bf Q}\, \hat{n}\cdot \boldsymbol{\sigma}_{\alpha \beta}.
\end{eqnarray}
Note that we have denoted the magnitudes of all the order parameters by  $\Phi_{\bf Q}$ for notational simplicity. 

\begin{figure}[htbp]
\begin{center}
\includegraphics[scale=0.5]{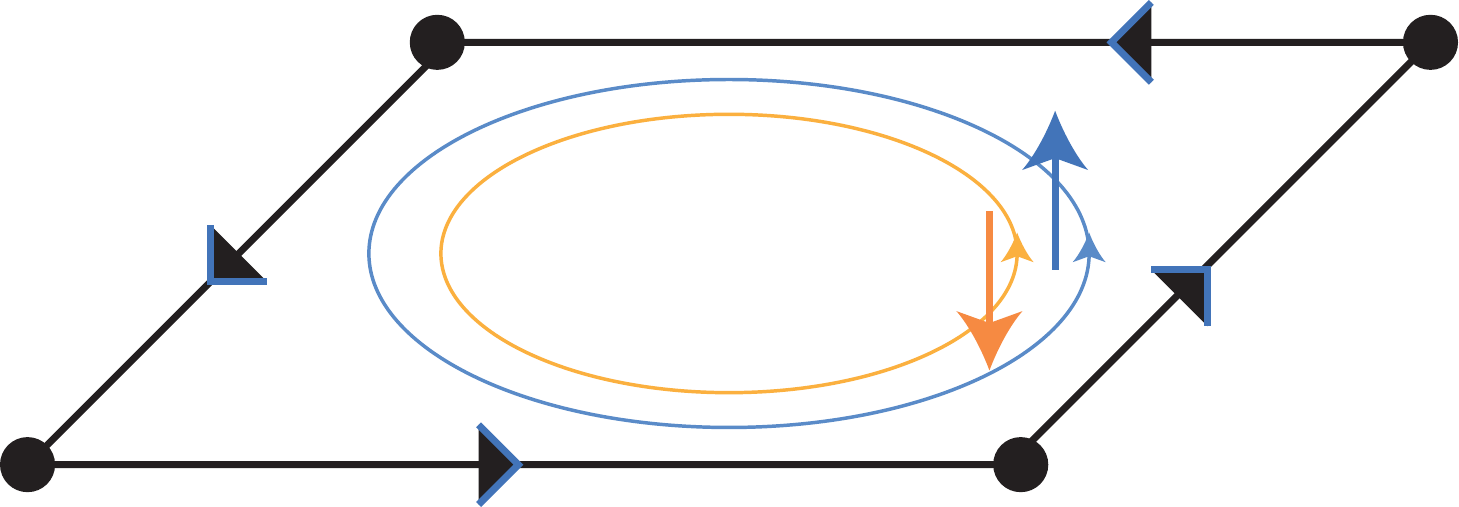}
\caption{(Color online) Circulating charge current for a single square plaquette corresponding to singlet DDW order parameter. Note that there is no circulating spin current.}
\label{fig:singlet}
\end{center}
\end{figure}

The sDDW modulates neither charge nor spin but represents staggered circulating charge currents, as shown in Fig.~\ref{fig:singlet}
In this case, the order parameter breaks lattice translational symmetry, time reversal, parity, and a rotation by $\pi/2$, but the product of any two symmetry operations is preserved. The order parameter is ``hidden'' because most external probes do not couple to currents or variations of kinetic energy. The order parameter corresponding to $S=0$, as in a sDDW,  does not respond to magnetic field directly, as in a paramagnet, but the quasiparticles energies are split  by the Zeeman effect. 

\begin{figure}[htbp]
\begin{center}
\includegraphics[scale=0.5]{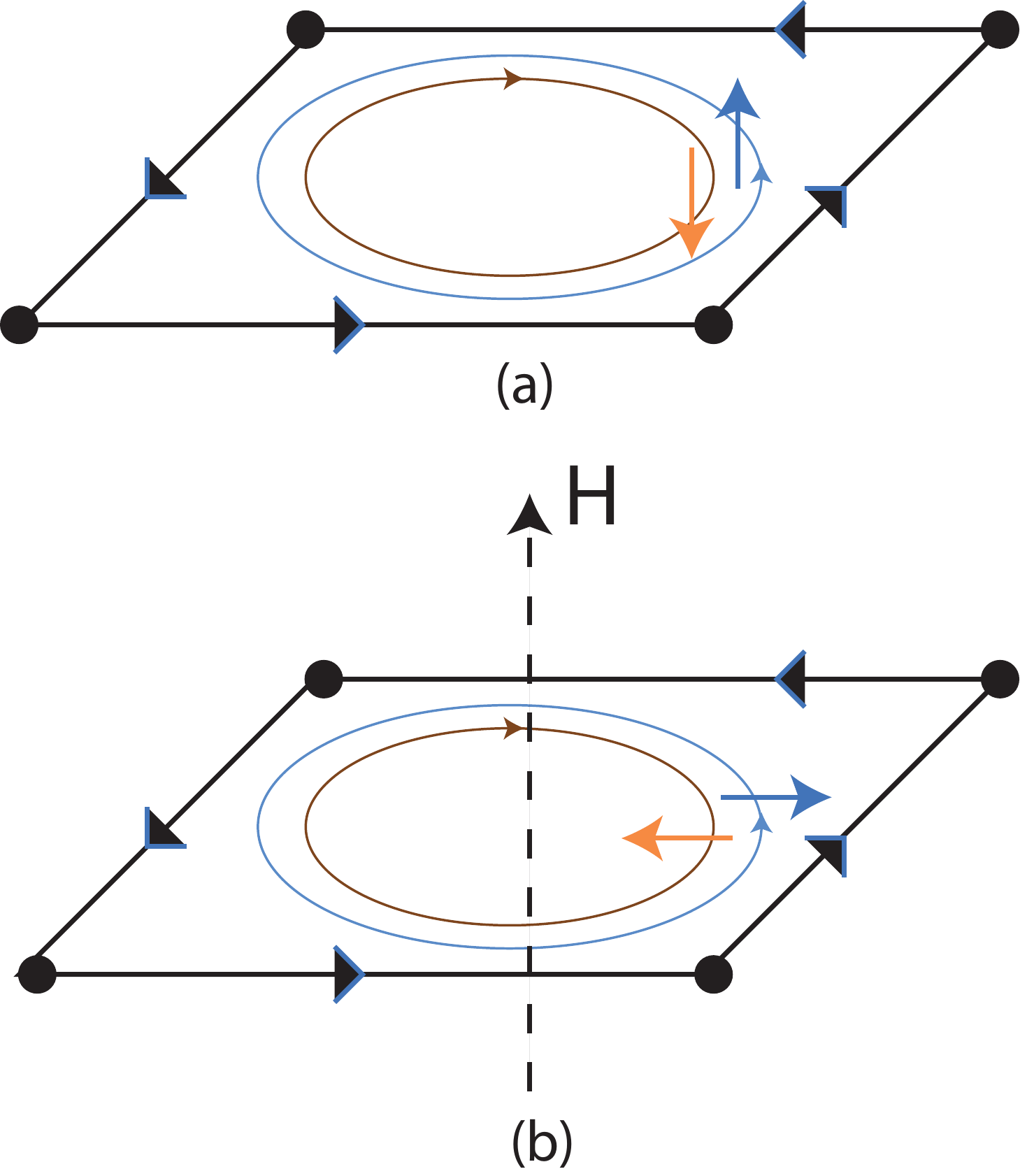}
\caption{(Color online) Spin-flop of triplet DDW in a magnetic field $H$ pictured for a single square CuO plaquette  with O atoms ommited. There is circulating spin current but no charge current. Here  (a) corresponds to the situation when $H=0$ and (b) to $H\ne 0$.}
\label{fig:spin-flop}
\end{center}
\end{figure}

The tDDW is further hidden because it is invariant under time reversal unlike SDW. Nonetheless both break spin rotational symmetry and lead to Goldstone modes. 
The tDDW, as well as the SDW, undergo a spin-flop transition in an arbitrary small magnetic field for zero spin-orbit interaction. If there is anisotropy in the spin space from spin-orbit coupling, there will be a non-zero threshold field beyond which the spin-flop transition will take place. In cuprates spin-orbit coupling is small enough that for high magnetic fields relevant for quantum oscillations, it is almost certain that the spins will be perpendicular to the applied magnetic field $H$, as shown in Fig.~\ref{fig:spin-flop}. There is a particularly nice way to bring out the similarities of tDDW and SDW, if we define the macroscopic order parameters by 
\begin{eqnarray}
{\bf y}&=&i\sum_{{\bf k}}f({\bf k})\sum_{\alpha,\beta}\boldsymbol{\sigma}_{\beta \alpha}\langle c^{\dagger}_{\alpha}({\bf k}+{\bf Q})
c_{\beta}({\bf k})\rangle,\\
{\bf m}&=&\sum_{{\bf k}}\sum_{\alpha,\beta}\boldsymbol{\sigma}_{\beta \alpha}\langle c^{\dagger}_{\alpha}({\bf k}+{\bf Q})
c_{\beta}({\bf k})\rangle.
\end{eqnarray}
Note the trace operation in the above definitions.
The tDDW and the spin flop in the presence of a magnetic field is shown in Fig.~\ref{fig:spin-flop}
\section{Spin Zeros}
In this section we shall calculate the spin zeros in quantum oscillations following the Lifhsitz-Kosevich (LK) formalism,~\cite{Abrikosov:1988} which we do not duplicate here. Only the relevant aspects pertaining to the reconstructed Fermi surfaces due to sDDW, tDDW and SDW will be described. In addition, we restrict ourselves to two dimensions without $c$-axis warping and bilayer splitting in  $\mathrm{YBa_{2}Cu_{3}O_{6+\delta}}$.~\cite{Garcia-Aldea:2010} This simplicity reveals the essential elements without unnecessary clutter. The extension to include the neglected effects are entirely straightforward. However,
we do need to specify the various parameters needed for the illustrative calculations, and these are summarized in the Appendix~\ref{appendix:A}. In this appendix we also provide the frequencies in the absence of the Zeeman field.
\subsection{Singlet}
The spectra of excitations for sDDW can be obtained from the following Hamiltonian
\begin{equation}
\mathbb{H}=\sum_{{\bf k}\in RBZ}\mathbf{\Psi }^{\dag }_{\bf k}\mathbb{A}_{\bf k}\mathbf{\Psi}_{\bf k}, 
\end{equation}
where
the reduced Brillouin
zone (RBZ) is bounded by $k_y \pm k_x = \pm\pi/a$ and 
\begin{equation}
\mathbb{A}=\left( 
\begin{array}{cccc}
\epsilon_{\uparrow,\bf k}& iW_{\bf k} & 0 & 0 \\ 
-iW_{\bf k} &\epsilon_{\uparrow,\bf k+Q} & 0 & 0 \\ 
0 & 0 & \epsilon_{\downarrow,\bf k} & iW_{\bf k} \\ 
0 & 0 & -iW_{\bf k} & \epsilon_{\downarrow,\bf k+Q}
\end{array}.
\right) 
\end{equation}
Here $\epsilon_{\uparrow,\bf{k}}=\epsilon_{\bf k}+\frac{g}{2}\mu_{B}H$ and $\epsilon_{\downarrow,\bf{k}}=\epsilon_{\bf k}-\frac{g}{2}\mu_{B}H$, where $\mu_{B}$ is the Bohr magneton. The four component spinor is given by $\Psi^{\dagger}_{\bf k}=(c_{\mathbf{k,}\uparrow }^{\dag },c_{\mathbf{k+Q,}\uparrow }^{\dag },c_{\mathbf{k,}\downarrow }^{\dag }, c_{\mathbf{k+Q,}\downarrow }^{\dag })$. Of course the up and the down spin sectors are decoupled, and the eigenvalues are
\begin{equation}
\label{eq:singlet}
\lambda_{\sigma,{\bf k}}^{s}=\frac{[\epsilon_{\bf k}+\epsilon_{\bf k+Q}]}{2}\pm \sqrt{\frac{[\epsilon_{\bf k}-\epsilon_{\bf k+Q}]^{2}}{4}+W_{\bf k}^{2}}+\frac{g}{2}\sigma\mu_{B}H, 
\end{equation}
which clearly shows Zeeman splitting as for a free spin.
In what follows,  to simplify the notation, we shall define:
\begin{eqnarray}
\epsilon_{1}&\equiv \frac{1}{2}[\epsilon_{\bf k}-\epsilon_{\bf k+Q}]\\
\epsilon_{2}&\equiv\frac{1}{2}[\epsilon_{\bf k}+\epsilon_{\bf k+Q}]
\end{eqnarray}
and drop the wave vector arguments when there is not any possibility of confusion.
For notational simplicity, we denote $\sigma=+1\equiv \uparrow$ and $\sigma=-1\equiv \downarrow$. 

The Zeeman term depends on the total magnetic field. It is included exactly within the Hamiltonian, because it is crucial in determining the sensitive interference of the amplitudes of quantum oscillations related to spin. The effect of the magnetic field on the orbital part is treated differently. First, its  effect on the DDW order parameter can be neglected,  as long as the system is deep inside the DDW phase; the fields relevant to quantum oscillations are energetically weak perturbations on the DDW gap.~\cite{Nguyen:2002} Clearly, close to a quantum phase transition where the DDW gap collapses, this will no longer be true---a situation that is not relevant to the experiments addressed here. Second,  in calculating quantum oscillations using LK formalism only the extremal Fermi surface areas and its various derivatives are needed, which can be calculated in the absence of the orbital part of the magnetic field. The oscillations arising form the extremal in-plane orbits are of course determined by only the normal component $H_{z}$; for simplicity we are ignoring  warping and bilayer splitting.  At the wave vector $(\pi/2a,\pi/2a)$, and at symmetry related points in the Brillouin zone, there are nodes. If the chemical potential, $\mu$, is placed at these nodes, the orbital diamagnetic susceptibility diverges  as $H\to 0$ in the absence of any interlayer coupling and at zero temperature, $T=0$.~\cite{Nersesyan:1991} However, the values of $\mu$ considered here are  too far from the nodal points for this to be relevant.

From the LK formalism   it is immediately obvious that the spin interference factor, $R_{s}$, within the LK formalism is 
\begin{equation}
R_{s}=\cos \left(\pi p \frac{m^{*}}{m}\frac{g}{2}\frac{1}{\cos \theta}\right),
\label{eq:Rs-singlet}
\end{equation}
where $\cos \theta$ is the angle between the magnetic field with respect to the normal and $p$ stands for the $p$-th harmonic.
Within our mean field theory $g=2$. The singlet DDW order parameter must exhibit spin zeros corresponding to elementary fermionic excitations of charge $e$, spin $1/2$, $g=2$, and the cyclotron  mass $m^{*}$ calculated from Eq.~\ref{eq:singlet}. Nonetheless,
the system is not a paramagnet because the spectra in Eq.~\ref{eq:Rs-singlet} correspond to a broken symmetry state far from a conventional paramagnet but with total spin $S=0$. Within Fermi liquid theory, with the reconstructed Fermi surface, Eq.~\ref{eq:Rs-singlet} should hold to all orders in perturbation theory. However, residual Fermi liquid  interactions between the quasiparticles can certainly lead to renormalization of $m^{*}$ and $g$. However, quite generally, within the Fermi liquid formalism, the residual interactions can only increase $m^{*}$ and $g$, assuming small spin orbit coupling, which appears to be the case; see Ref.~\onlinecite{Ramshaw:2010} and reference therein. Thus the case for the existence of spin zeros will be stronger if the Fermi liquid corrections are taken into account. 

\subsection{Triplet}
\begin{figure}[htbp]
\begin{center}
\includegraphics[scale=0.5]{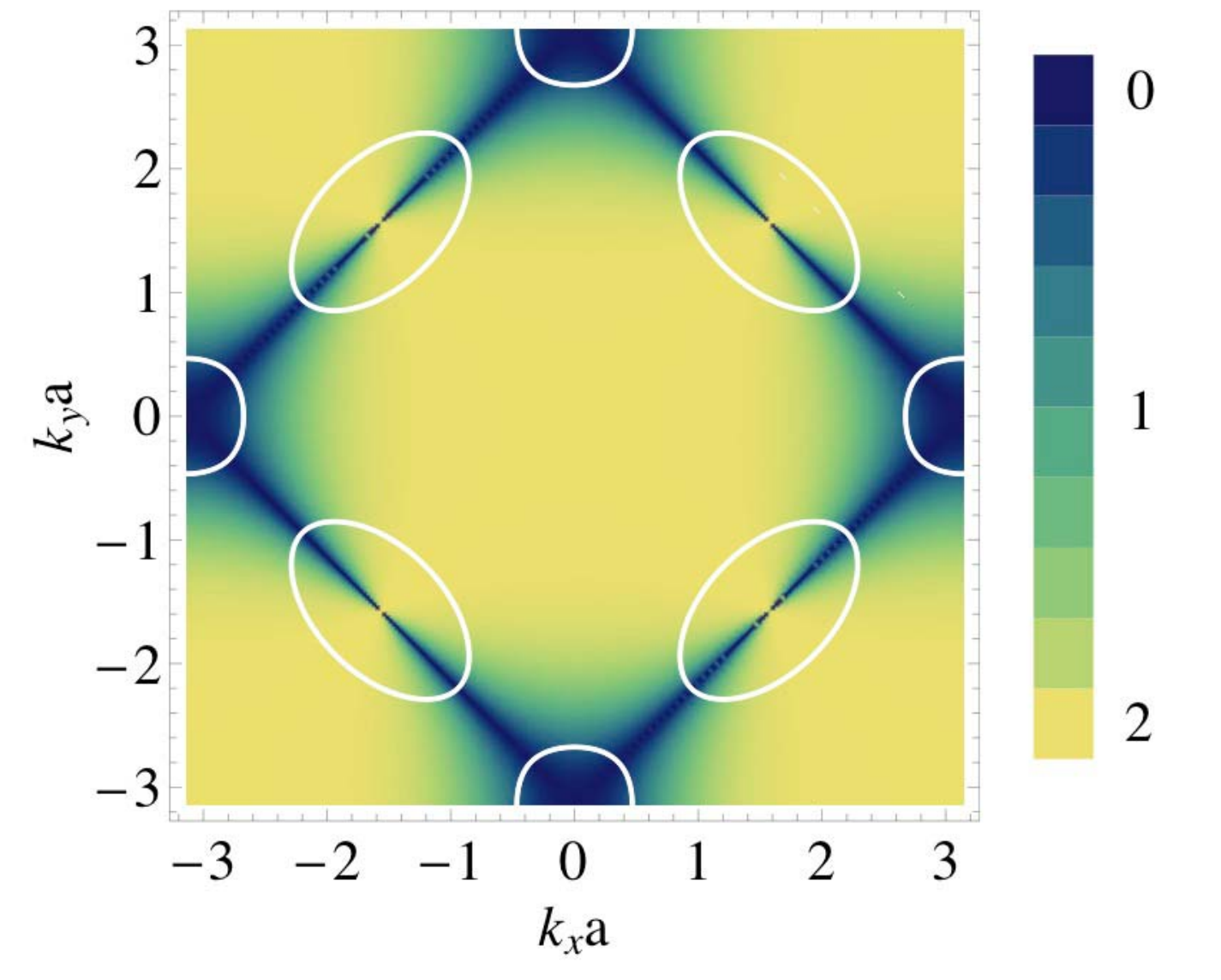}
\caption{(Color online) A map of $g_{\text{eff}}({\bf k})$ for tDDW. Note that it is mostly 2 for most of the hole pocket and mostly zero for the electron pocket. This effective g-factor will split the hole and the electron pockets into two close pockets depending on the magnetic field. The actual splitting due to Zeeman coupling is not visible on the scale of the plot.}
\label{fig:g-factor}
\end{center}
\end{figure}

\begin{figure}[htbp]
\begin{center}
\includegraphics[scale=0.35]{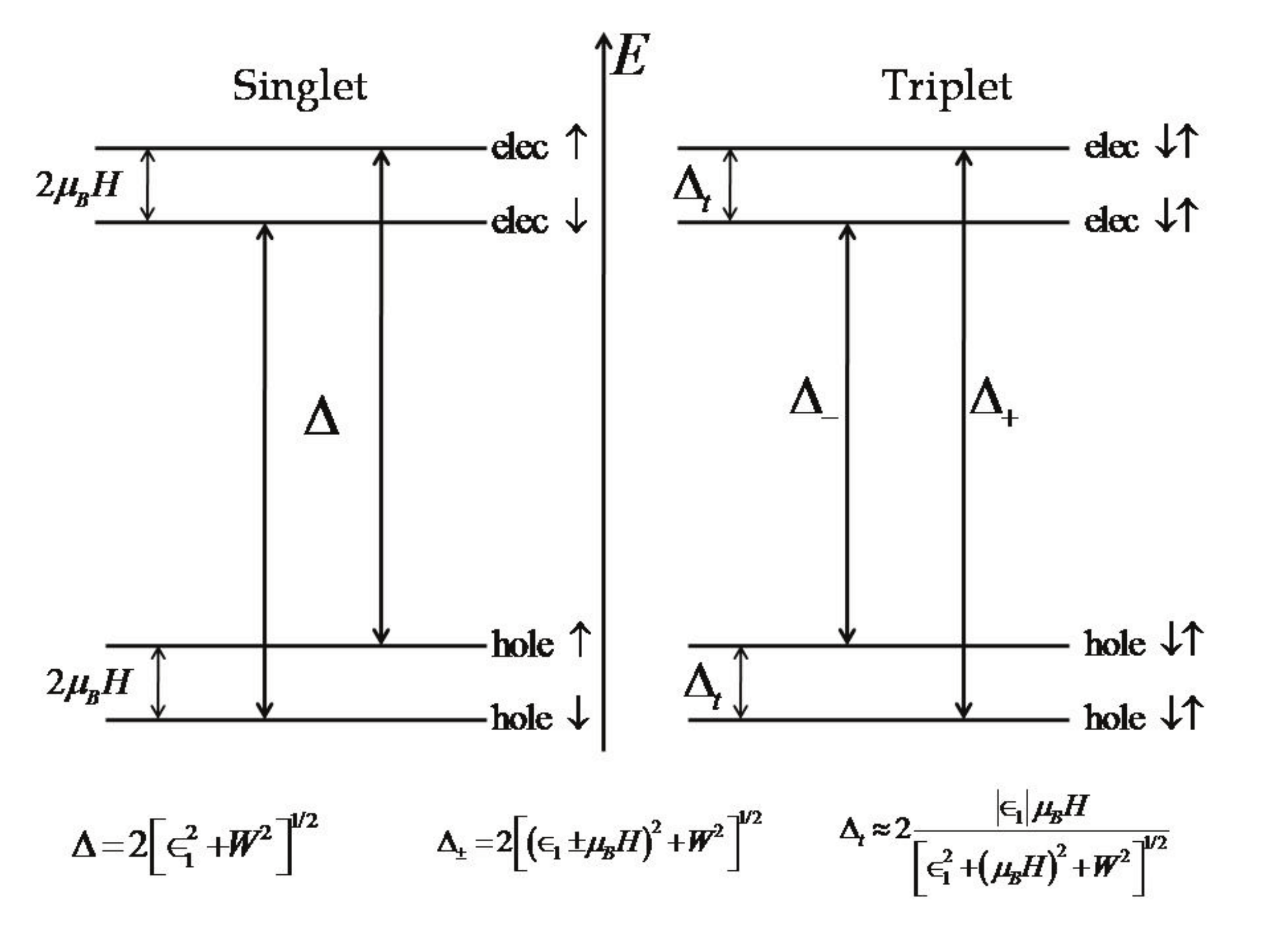}
\caption{The splitting of the levels in a magnetic field. Note that in the triplet case we cannot identify the levels with a definite spin quantum number. The hole and the electron pockets are labelled ``hole'' and ``elec''.}
\label{fig:e-level}
\end{center}
\end{figure}

In the triplet  case the spin orientation is chosen to be perpendicular to the direction of the applied field  because of the spin flop.  Thus, with no loss of generality,  we can choose $H_{z}=H \cos {\theta}$, $H_{x}=$ $H_{y}=0$ and $n_{x}=1$, $
n_{y}=n_{z}=0$. Then  the matrix $\mathbb{A}$ is
\begin{equation}
\mathbb{A}=\left( 
\begin{array}{cccc}
\epsilon_{\uparrow,\bf k} & 0 & 0 & iW \\ 
0 & \epsilon_{\uparrow,\bf k+Q}& -iW & 0 \\ 
0 & iW & \epsilon_{\downarrow,\bf k} & 0 \\ 
-iW & 0 & 0 &\epsilon_{\downarrow,\bf k+Q}
\end{array}
\right) 
\label{eq:t-matrix}
\end{equation}
The eigenvalues are
\begin{equation}
\label{eq:triplet1}
\lambda_{\sigma,{\bf k}}^{t}=\epsilon_{2}\pm \sqrt{(\epsilon_{1}+\frac{g}{2}\sigma\mu_{B}H)^{2}+W^{2}}
\end{equation}
The four eigenvectors are various linear superpositions involving the coherence factors of the original fermion operators and  do not have definite spin unlike the singlet case. For arbitrary $\bf k$ the mixing  depends on the $\bf k$-space orbit, and this gives rise to a dynamically generated spin-orbit effect, as can be seen from the quasiparticle spectra, if we expand in powers of the magnetic field. The differences between the energies of the two close energy levels, either the  holes or the electrons, are given by
\begin{eqnarray}
\Delta_{h}=\Delta_{e}&\approx& g\mu_{B}H\frac{|\epsilon_{1}|}{\sqrt{\epsilon_{1}+W^{2}+(g\mu_{B}H/2)^{2}}},\\
&\equiv&g_{\text{eff}}({\bf k})\mu_{B}H,
\end{eqnarray}
that is, by an effective $g$-factor. A typical plot of $g_{\text{eff}}$ is shown in Fig.~\ref{fig:g-factor}.
Another way to picture the distinction between the singlet and the triplet cases is the schematic energy level diagram shown in Fig.~\ref{fig:e-level}.

\subsubsection{Effective $g$-factor}
 Let us  focus on
the electron pockets; the results for the hole pockets follow identically. Let
\begin{equation}
\varepsilon_{\pm}=\epsilon _{2}+\left[ \left( \epsilon _{1}\pm\frac{g}{2} \mu
_{B}H\right) ^{2}+W^{2}\right] ^{1/2} 
\label{eq:triplet}
\end{equation}
 To calculate the interference factor due to the two close electron pockets within the LK formalism, we must compute the 
 sum
\begin{equation}
\begin{split}
\sum_{j=\pm}e^{2\pi ipn_{mj}}&= e^{2\pi ipn_{m+}\left( \mu \right) }e^{2\pi
ip\left( \frac{\partial n_{m+}}{\partial \varepsilon _{+}}\right) _{\mu
}\left( \varepsilon _{+}-\mu \right) }\\
&+e^{2\pi ipn_{m-}\left( \mu \right)
}e^{2\pi ip\left( \frac{\partial \bar{n}_{m-}}{\partial \bar{\varepsilon} _{-}}\right)
_{\mu }\left( \varepsilon _{-}-\mu \right) }. 
\end{split}
\end{equation}
corresponding to the Landau level index $n_{m\pm}$ corresponding to the extremal Fermi surface orbits.
Since the splitting is small for experimentally relevant $H$, it is a good approximation to assume that
$n_{m+}\left( \mu \right)\approx n_{m-}\left( \mu \right)\approx {\bar n}_{m}\left( \mu
\right)$. Similarly,
\begin{equation}
\left( \frac{\partial n_{m+}}{\partial \varepsilon _{+}}\right) _{\mu
}\approx\left( \frac{\partial n_{m-}}{\partial \varepsilon _{-}}\right) _{\mu
}\approx\left( \frac{\partial \bar{n}_{m}}{\partial \bar{\varepsilon}}\right) _{\mu },
\end{equation}
where $\bar{n}_{m}=(n_{m+}+n_{m-})/2$, $\varepsilon _{+} = \bar{\varepsilon}+\Delta \varepsilon$, and $\varepsilon _{-} =\bar{\varepsilon}-\Delta \varepsilon$.
Thus,
\begin{equation}
\begin{split}
\sum_{j}e^{2\pi ipn_{mj}} 
&\approx e^{2\pi ip\bar{n}_{m}\left(\mu \right) }\exp\left( 2\pi ip\left( \frac{\partial
\bar{n}_{m}}{\partial \bar{\varepsilon}}\right) _{\mu }\left( \bar{\varepsilon}
-\mu \right)\right)\\
&\times \cos \left( 2\pi p\left( \frac{\partial \bar{n}_{m}}{\partial \bar{
\varepsilon}}\right) _{\mu }\Delta \varepsilon \right) .
\end{split}
\end{equation}
We can set $\left( \bar{\varepsilon}-\mu \right) \approx 0$ and the identification of $\bar{\varepsilon}$ and $\Delta \epsilon$ follows trivially from Eq.~\ref{eq:triplet}. 
Then the interference factor is 
\begin{equation}
R_{s}=\cos \left( \pi p\frac{m^{\ast }}{m_{e}}\frac{g_{\text{eff}}({\bf k})}{2}\frac{1}{\cos \theta}\right) 
\label{eq:Rs-triplet}
\end{equation}
As the $g_{\text{eff}}$ is $\mathbf{k}$-dependent we average
over the extremal orbits shown in Fig.~ \ref{fig:g-factor}. 
\subsubsection{Filling of the Landau levels}
There is another method to obtain the interference factor that serves as a consistency check, especially because we have made approximations. In this approach we focus
on the filling of the Landau levels. 
We can define the average  $\bar{n}=\frac{1}{2}
\left( n_{+}+n_{-}\right) $ and the difference $\Delta n=\frac{1}{2}\left(
n_{+}-n_{-}\right) $. Where the plus sign makes reference to the pocket that
is bigger and the minus sign to the pocket that is smaller. It does not
matter if the difference is due to the Zeeman spin splitting, as  for the singlet
case, or the splitting  in the spectra in the triplet case. If we extend this
notation to the extremal areas $m$, $\bar{n}
_{m}=\frac{1}{2}\left( n_{m+}+n_{m-}\right) $ and the difference $\Delta
n_{m}=\frac{1}{2}\left( n_{m+}-n_{m-}\right) $
Again taking advantage of small splitting, we can show that 
in the LK formula
\begin{equation}
\sum_{j=\pm}e^{2\pi i p n_{mj}} =
e^{2\pi i p\bar{n}_{m}(\mu)}\cos \left( 2\pi p\Delta n_{m}\right) 
\end{equation}
Now the dependences on the $g_{\text{eff}}$, cyclotron masses, and the tilt
angle are contained in $\Delta n_{m}.$ If we compare with the previous
results we obtained for the cosine factor we get
\begin{equation}
\cos \left( \pi p\frac{g_{s,t}}{2}\frac{m^{\ast }}{m_{e}}\frac{1}{\cos \theta }
\right) =\cos \left( 2\pi p\Delta n_{m}\right) . 
\end{equation}
valid for both singlet and triplet. If $\Delta n_{m}=0$ the
argument vanishes and the cosine is unity; there will be no spin zeros at all.
We can deduce what is the effective $g$-factor for the triplet if we know
the result for the singlet: 
$\left( \Delta n_{m}\right) _{t}/\left( \Delta n_{m}\right)
_{s}=g_{t}m^{*}_{t}/g_{s}m^{*}_{s}$.
Assuming that the effective cyclotron masses are approximately the same (see Table~\ref{effmass1}) for
sDDW and tDDW, we get
\begin{equation}
g_{t}\approx g_{s}\frac{\left( \Delta n_{m}\right) _{t}}{\left(
\Delta n_{m}\right) _{s}}. 
\label{eq:gt-Landau}
\end{equation}
\begin{table}[htdp]
\caption{Calculated cyclotron masses in units of the free electron mass for sDDW, tDDW, and SDW order parameters. e-p and h-p correspond respectively to electron and hole pockets, and the two rows  to the two distinct Fermi surfaces. The errors are less than one part in 1000. The first row corresponds to the inner pocket and the second to the outer pocket.}
\begin{center}
\begin{tabular}{|c|c|c|c|c|c|}
    \hline
    \hline
   sDDW& sDDW & tDDW & tDDW &  SDW &  SDW \\
   e-p&h-p&e-p&h-p&e-p&h-p\\
    \hline
    \hline
    1.55 & 0.95 & 1.56 & 0.96 & 1.27 &1.12 \\
    1.55 & 0.97 & 1.54 & 0.96 & 1.28 & 1.13\\
    \hline
    \hline
\end{tabular}
\end{center}
\label{effmass1}
\end{table}
\begin{table}[htdp]
\caption{The effective $g$-factors. The computations were carried out for  40 Tesla  field for illustrative purposes. The Method 1 is based on $g_{\text {eff}}({\bf k})$ in Eq.~\ref{eq:Rs-triplet} and the  Method 2 is based on Eq.~\ref{eq:gt-Landau}. A reasonable estimate of the errors may be taken to be the difference of the results between the two methods.}
\begin{center}
\begin{tabular}{|c|c|c|c|c|}
\hline
\hline
$g_{\text{eff}}$ & tDDW &tDDW & SDW & SDW \\
&e-p&h-p&e-p&h-p\\
\hline
\hline
Method 1&0.32 & 1.65 & 0.45 & 1.42 \\
Method 2&0.31 & 1.65 &0.40 & 1.42\\
\hline
\hline
\end{tabular}
\end{center}
\label{Table2}
\end{table}
\begin{figure}[htb]
\begin{center}
\includegraphics[scale=0.5]{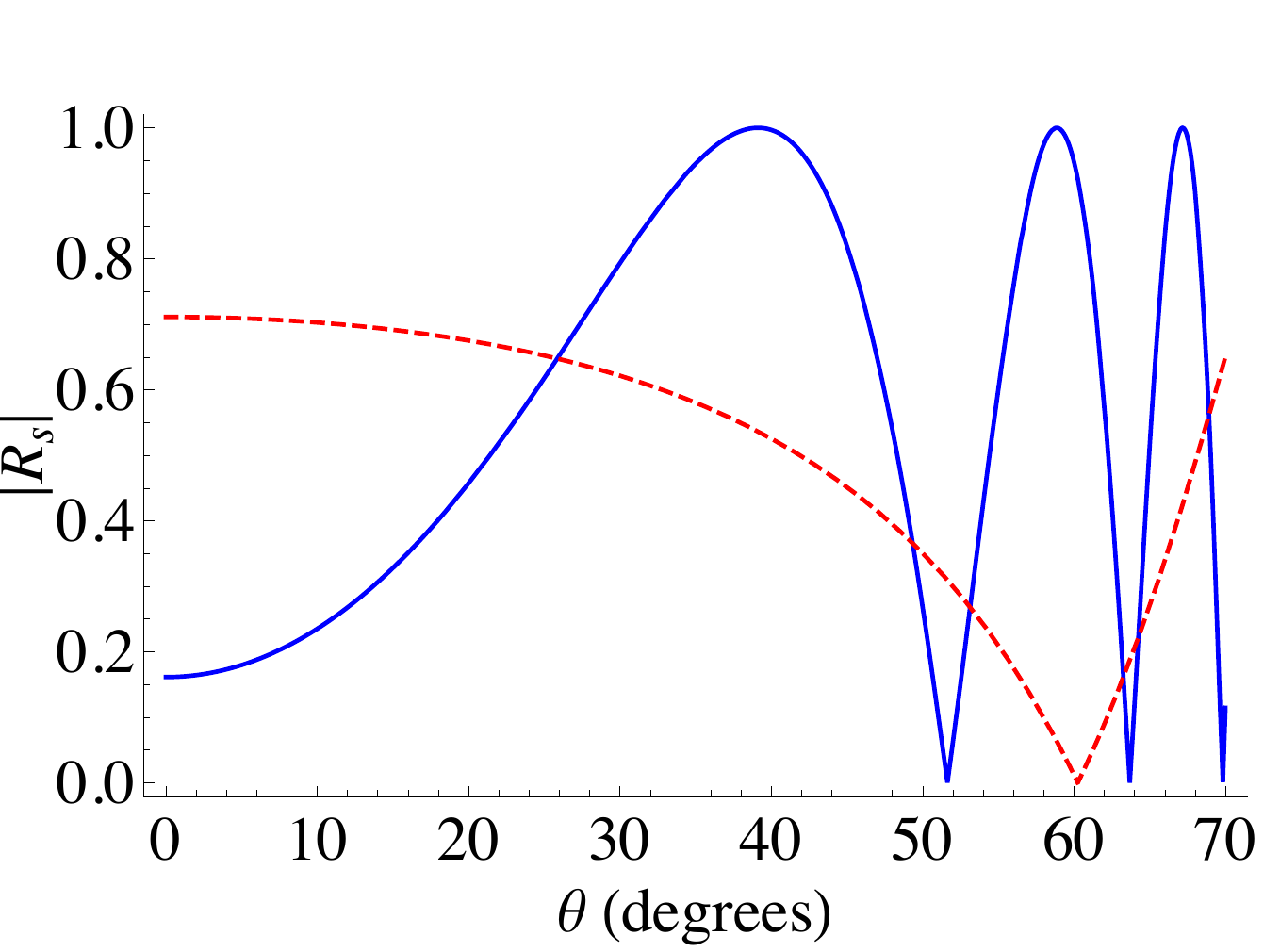}
\end{center}
\caption{(Color online) $|R_{s}|$ as a function of the angle of the tilted magnetic field for the electron pocket and for $p=1$. The solid line corresponds to sDDW and the dashed the tDDW.}
\label{fig:Rse}
\end{figure}

\begin{figure}[htb]
\begin{center}
\includegraphics[scale=0.5]{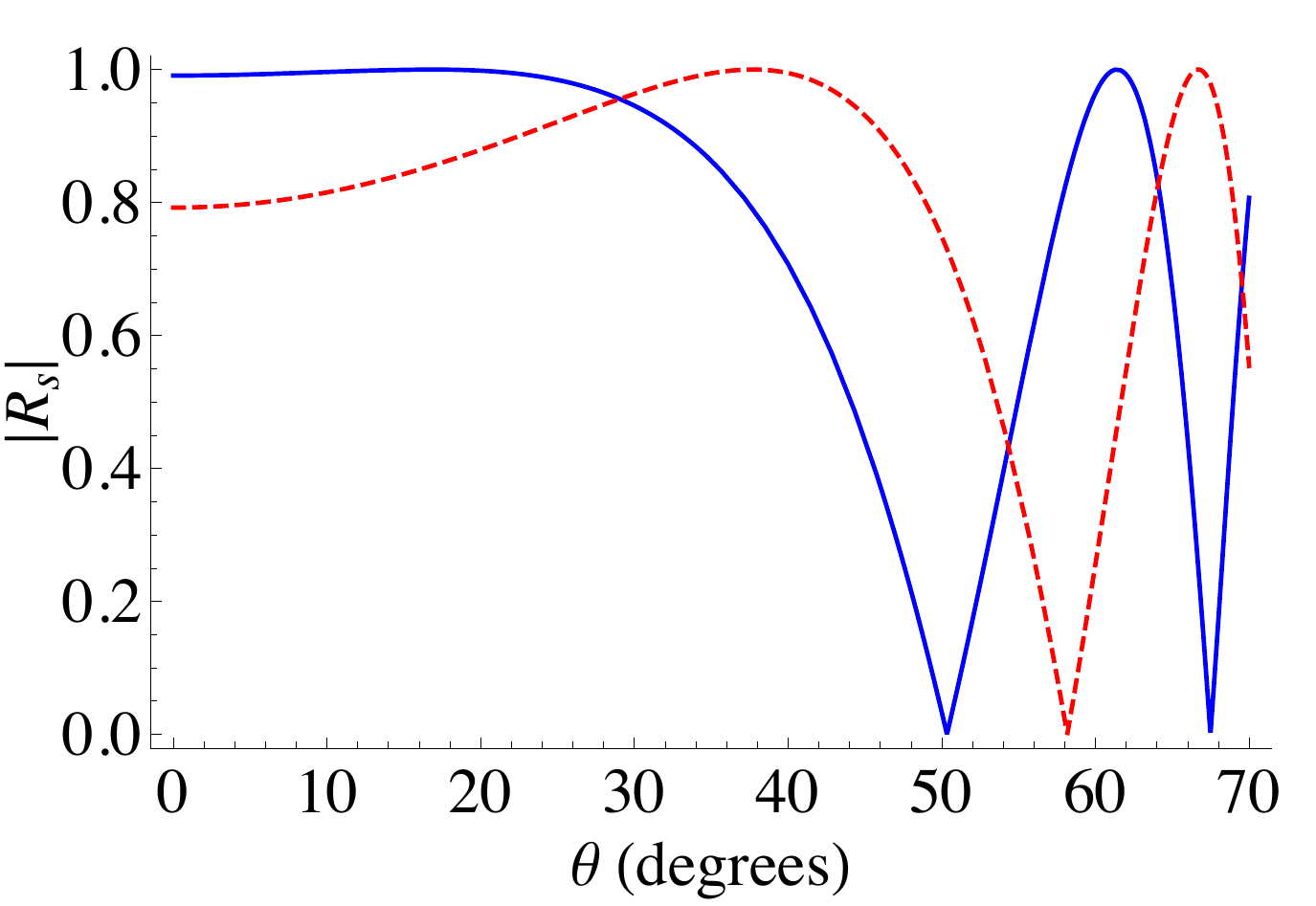}
\end{center}
\caption{(Color online)$|R_{s}|$ as a function of the angle of the tilted magnetic field for the hole pocket and $p=1$.The solid line corresponds to sDDW and the dashed the tDDW.}
\label{fig:Rsh}
\end{figure}
It is easy to see that that $n_{m}$ is proportional to the extremal Fermi surface areas.
For our purposes we can set $g_{s}=2$ to compute $g_{t}$ from
the  integration of the 
Fermi surface areas.
The results are shown below in Table~\ref{Table2}. The absolute value of $R_{s}$, the factor that is responsible for the spin zeros is shown for the electron pocket, as a function of the tilt of the magnetic field with respect to the normal  in Fig.~\ref{fig:Rse}. In the singlet case we have chosen the free electron value of $g_{s}=2$. In the triplet case the calculated effective $g_{t}=0.32$ is used for the electron pocket.   We can see that for the singlet a spin zero is found close to $50^{\circ}$, while the first spin zero for the triplet occurs above $60^{\circ}$. Similar results are shown for the hole pocket in Fig.~\ref{fig:Rsh} using $g_{t}=1.65$.
\subsection{Commensurate SDW}
\begin{figure}[htb]
\begin{center}
\includegraphics[scale=0.5]{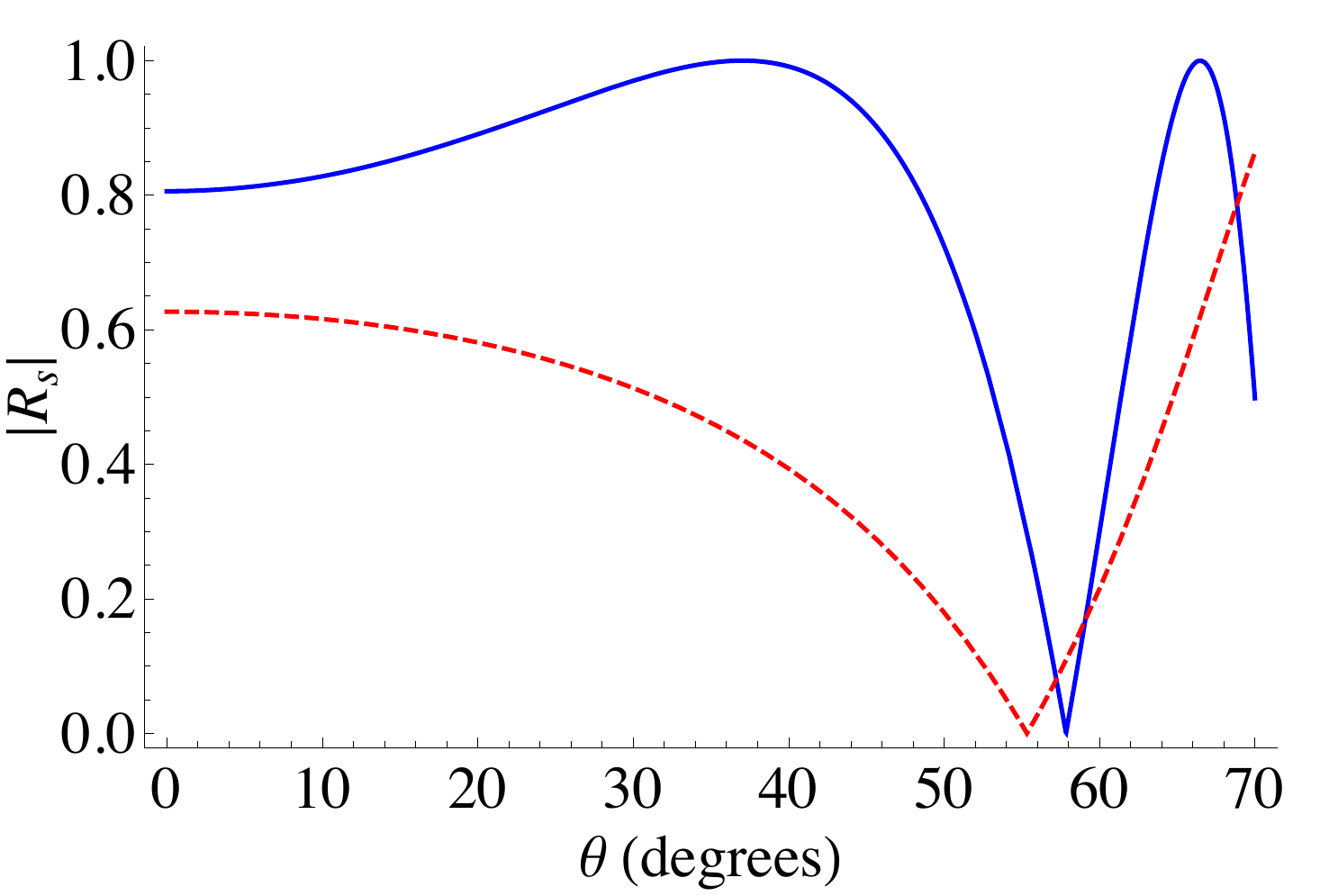}
\end{center}
\caption{(Color online) $|R_{s}|$ as a function of the angle of the tilted magnetic field for SDW and $p=1$. The solid line in this figure represents the hole pocket and the dashed line the electron pocket.}
\label{fig:SDW}
\end{figure}

A commensurate SDW in our scheme is an orbital $s$-wave order parameter for which the calculation proceeds identically except that the $f({\bf k})=1$. The results are shown in Fig.~\ref{fig:SDW}. 

\section{Conclusion}

The rigorous statement we can make is that a singlet order parameter will definitely lead to Zeeman splitting (see Eq.~\ref{eq:singlet}) and therefore spin zeros (recall the discussion of the g-factor above). A triplet order parameter, be it SDW or triplet-DDW, does not exhibit Zeeman splitting (see Eq.~\ref{eq:triplet1}). It does lead to some dependence of the amplitude of oscillations on the tilt in a non-universal manner, and may or may not lead to spin zeros within a reasonable angular range, such that the system is not driven into the superconducting state, which may lead to the loss of the amplitude of the oscillations. In other words, the experiments in Ref.~\onlinecite{Ramshaw:2010} are consistent with a singlet order parameter. 

It is remarkable that once the conventional band parameters are adopted, and the oscillation frequency is adjusted close to the experimentally observed one at 526 T for the electron pocket (for the other frequency, effects such as bilayer splitting and c-axis warping, must be taken into account), the oscillation amplitude as a function of the tilt angle for the singlet order parameter is very close to the measured amplitude in Ref.~\onlinecite{Ramshaw:2010}. As one can see from the plots, the triplet order parameter behaves very differently.

The spin zeros determine only the product $g m^{*}/m_{e}$, which calculated  in the case of sDDW for the electron pocket is 3.1, as opposed to 3.2 in Ref.~\onlinecite{Ramshaw:2010}. For the hole pocket, in contrast, it is calculated to be 1.92. Unfortunately there does not appear to be any evidence  of the hole pockets in experiments, about which we have commented elsewhere.~\cite{Jia:2009}  

A useful insight into tDDW can be obtained from the coherence factors corresponding to the  four eigenvectors of the matrix in Eq.~\ref{eq:t-matrix}. These are 
\begin{eqnarray*}
\gamma_{1,\bf k} &=&v_{1,\bf k}c_{\mathbf{k,}\uparrow }+u_{1,\bf k}c_{\mathbf{k+Q,}\downarrow }, \\
\gamma_{2,\bf k} &=&v_{2,\bf k}c_{\mathbf{k+Q,}\uparrow }+u_{2,\bf k}c_{\mathbf{k,}\downarrow }, \\
\gamma_{3,\bf k} &=&v_{3,\bf k}c_{\mathbf{k+Q,}\uparrow }+u_{3,\bf k}c_{\mathbf{k,}\downarrow },\\
\gamma_{4,\bf k} &=&v_{4,\bf k}c_{\mathbf{k,}\uparrow }+u_{4,\bf k}c_{\mathbf{k+Q,}\downarrow },
\end{eqnarray*}
where the coherence factors are
\begin{equation}
\left. 
\begin{array}{c}
\left\vert u_{1,\bf k}\right\vert ^{2}=\left\vert v_{4,\bf k}\right\vert ^{2} \\ 
\left\vert v_{1.\bf k}\right\vert ^{2}=\left\vert u_{4,\bf k}\right\vert ^{2}%
\end{array}%
\right\} =\frac{1}{2}\left( 1\pm \frac{\Delta\epsilon _{\bf{k},+}}{E_{\bf{k},+}}\right) ,
\end{equation}
\begin{equation}
\left. 
\begin{array}{c}
\left\vert u_{2,\bf k}\right\vert ^{2}=\left\vert v_{3,\bf k}\right\vert ^{2} \\ 
\left\vert v_{2,\bf k}\right\vert ^{2}=\left\vert u_{3,\bf k}\right\vert ^{2}%
\end{array}%
\right\} =\frac{1}{2}\left( 1\pm \frac{\Delta\epsilon _{{\bf k},-}}{E_{\bf{k},-}}\right) 
\end{equation}
and  
\begin{eqnarray}
E_{{\bf k},\sigma} &=&\left[ \left( \epsilon _{1}+\sigma\frac{g}{2} \mu
_{B}H\right) ^{2}+W^{2}\right] ^{1/2},  \\
\Delta\epsilon _{{\bf k},\sigma} &=&\epsilon_{1}+\frac{g}{2}\sigma\mu _{B}H .
\end{eqnarray}
The coherence factors reflect the fact that the quasiparticles do not have a definite spin. For arbitrary $\bf k$ the mixing is not of equal amplitude.

Another amusing observation is that as far as the product $g m^{*}/m_{e}$ is concerned both tDDW and SDW yield essentially identical answers even though the individual values for the effective $g$-factors and the $m^{*}$ are different. For the electron pocket this product is 0.5 for tDDW and 0.51 for SDW. Similarly, for the hole pocket it is 1.59 for tDDW and 1.60 for SDW.

One might wonder if there are other possibilities of a singlet order parameter that could be a candidate broken symmetry state. In principle one cannot rule out CDW order, although such an order parameter would not have gone unnoticed in many other direct experiments. Thus, the existence of spin zeros may be consistent with sDDW,  whose direct  observation by its very nature may be considerably hidden.

\section*{Acknowledgments}
We thank  Doug Bonn, Cyril Proust, S. Sebastian, and especially Brad Ramshaw for keeping us informed of their experiments.   This work is supported by NSF under the Grants DMR-0705092 and DMR-1004520. DGA acknowledges financial support to Fundacion Cajamadrid through its
grant program and from the Spanish MCInn through the research project ref.
FIS2007-65702-C02-02.

{\em Note added:} After our work was completed we learned of two interesting papers that appear to be complementary. In one, the problem has been addressed from the perspective of stripe physics (arXiv: 1007.1047v1) and in the other antiferromagnets are studied (arXiv:1006.0167v1).

\appendix
\section{\label{appendix:A}Parameters}
We use a common band structure,~\cite{Andersen:1995,Pavarini:2001}
\begin{equation}
\begin{split}
\epsilon_{\bf k}&=-2t(\cos{k_x}a+\cos{k_y}a)+4t'\cos{k_x}a\cos{k_ya}\\&-2t''(\cos{2k_x}a+\cos{2k_y}a),
\end{split}
\end{equation}
where $t=0.154\;  eV, \;  t'=0.32t, \; t''=0.5t'$. For sDDW and tDDW we choose~\cite{Nayak:2000,Chakravarty:2001,Nersesyan:1991}
\begin{equation}
W_{\bf k}=\frac{W_{0}}{2}(\cos k_{x}a - \cos k_{y}a),
\end{equation}
with $W_{0}=0.9 t$. The chemical potential $\mu=-0.775t$ then results in a hole doping of approximately $10.7\%$. For SDW the choices were  $W_{\bf k}=W_{0}=0.675 t$ and $\mu = - 0.93 t$, yielding a hole doping of approximately $10.6\%$.
The frequencies corresponding to the Fermi surface areas in the absence of Zeeman splitting are (a) sDDW: 513 Tesla for the electron pocket and 1005 Tesla for the hole pocket; (b) tDDW: 513 Tesla for the electron pocket and 1004 Tesla for the hole pocket; (c) SDW: 539 Tesla for electron pocket and 1012 for the hole pocket.

\end{document}